\documentstyle[amssymb,12pt]{article}

\begin{document}

\title{Saint-Venant's principle in dynamical porous thermoelastic media with memory
for heat flux}
\author{Gerardo Iovane, \quad Francesca Passarella \thanks{
E-mail address: iovane@diima.unisa.it, passarella@diima.unisa.it} \and
{\small Department of Information Engineering and Applied Mathematics
(DIIMA),}}
\date{{\small University of Salerno, 84084 Fisciano (Sa), Italy}}
\maketitle

\begin{quote}
{\bf Abstract} -- {\small In the present paper, we study a linear
thermoelastic porous material with a constitutive equation for heat flux
with memory. An approximated theory of thermodynamics is presented for this
model and a maximal pseudo free energy is determined. We use this energy to
study the spatial behaviour of the thermodynamic processes in porous
materials. We obtain the domain of influence theorem and establish the
spatial decay estimates inside of the domain of influence. Further, we prove
a uniqueness theorem valid for finite or infinite body. The body is free of
any kind of a priori assumptions concerning the behaviour of solutions at
infinity. }
\end{quote}

\section{Introduction}

The Saint--Venant's principle has a central role in more theoretical and
applied questions of elasticity. An important review of research on the
spatial behaviour of solutions for statical and dynamical problems was given
by Horgan and Knowles \cite{1983} and Horgan \cite{1989,1996}. Relevant
information on the spatial behaviour of the solutions for the dynamical
problems of elasticity are given by the domain of influence theorem as it is
presented by Gurtin \cite{6}. A further study in this connection was made by
Chiri\c{t}\u{a} \cite{7} for the linear theory of thermoelasticity, where a
bounded solid is subjected to the action of nonzero boundary loads only on
the end face.

Recently, Fabrizio, Lazzari and Munoz Rivera \cite{Falamu} have studied a
linear thermoelastic material which exhibits a constitutive equation for
heat flux with memory. Some existence, uniqueness and asymptotic theorems
have been established in connection with this approach. Further, Chiri\c{t}%
\u{a} and Lazzari \cite{5} have studied the spatial behaviour of the
thermoelastic processes. The hereditary effects are taken into account in a
dissipative boundary condition by Ciarletta \cite{ciarletta} and Bartilomo
and Passarella \cite{barpass}.

On the other hand, Ie\c{s}an \cite{iesan} has developed a linear theory of
thermoelastic materials with voids, by generalizing some ideas of Cowin and
Nunziato \cite{1}. In this theory the bulk density is written as the product
of two fields, the matrix material density field and the volume fraction
field. This representation introduces an additional degree of kinematic
freedom and it is compatible with the theory of granular materials developed
by Goodman and Cowin \cite{2}. The spatial and temporal behaviour in linear
thermoelasticity of materials with voids has been studied by Chiri\c{t}\u{a}
and Scalia in \cite{ScStan}. Further,\ it is shown by Iovane and Passsarella
in \cite{Passarella} the spatial exponential decay for the boundary--final
value problem associated to elastic porous materials with a memory effect
for the intrinsic equilibrated body forces.

In the present paper, we study a linear thermoelastic porous material which
exhibits effects of fading memory in the constitutive equation for the heat
flux. Such model exhibits finite speeds for the propagation of thermal
disturbances and hence leads to accurate modeling of the transient thermal
behaviour. By applying the thermodynamical laws, we can prove the existence
of a pseudofree energy potential such that the internal dissipation vanishes
despite of the dissipative character of the model. Then, we use this energy
for studying the spatial behaviour of thermodynamical processes.

The paper is organized as follows. In Section 2, we present the
thermodynamical restrictions imposed by the thermodynamical laws on the
model in question; then, we introduce a maximal pseudofree energy. In
Section 3, we deduce\ a result for describing the domain of influence by
using an appropriate time-weighted surface measure. This domain at instant $%
t\in \lbrack 0,\ T]$ is the set $D_{\zeta t}$ of all points of continuum for
which $r\leq \zeta t$, where $r$ represents the distance from the bounded
support ${\hat{D}}_{T}$ of the (external) given data in the time interval $%
[0,\ T]$ and $\zeta $ is a constant characteristic to the thermodynamic
coefficients. We obtain, into the inner of the domain of influence, a
spatial decay of exponential type. The decay rate, characterized by a factor
independent of time, is $\displaystyle\exp (-\frac{\sigma }{\zeta }r)$,
where $\sigma $ is a positive parameter independent of time and defining the
time-weighted surface power measure. In Section 4, we prove that, into the
inner of the domain of influence\ $D_{\zeta t},$ an energetic measure
associated with the thermodynamic process tends to zero with a decay rate
equal to $\left( 1-\frac{r}{\zeta t}\right) .\ $

\section{Basic equations. Thermodynamic restrictions}

We consider a porous thermoelastic body that occupies a (regular) region $B$
of the physical space ${\rm I}\!{\rm R}^{3}$ in an assigned reference
configuration. By identifying ${\rm I}\!{\rm R}^{3}$ with the associated
vector space, an orthonormal system of reference is introduced, so that
vectors and tensors will have components denoted by Latin subscripts ranging
over 1, 2, 3. The letters in boldface stand for the tensor ${\bf L}$ of an
order $p\geq 1,$ and $L_{ij{\it .\ .\ .\ }k}$ ($p$ subscripts) are the
components of the tensor ${\bf L}$. Summation over repeated subscripts and
other typical conventions for differential operations are implied, such as a
superposed dot or a comma followed by a subscript to denote partial
derivative with respect to time or with respect to the corresponding
coordinate.

We denote ${\bf U}\equiv \{{\bf u},\ \varphi ,\ \theta \},\ $where ${\bf u}$
is the displacement vector fields, $\varphi $ is the change in volume
fraction starting from the reference configuration, $\theta $ is the
temperature variation from the uniform reference temperature $\theta
_{0}(>0) $. Further, $\rho $ is the bulk mass density and $\stackrel{}{\chi }
$ is the equilibrated inertia in the reference state.

We restrict our attention to the linear theory of thermoelasticity in which
the effect of fading memory is contained in the thermal gradient ${\bf %
\triangledown }\theta {\it .}$ The local balance equations become \cite
{iesan} for the present problem\
\begin{equation}
\begin{array}{l}
S_{ji,j}+\rho f_{i}=\rho \ddot{u}_{i}, \\[4mm]
h_{j,j}+g+\rho \ell =\rho \chi \ddot{\varphi}, \\[4mm]
-q_{j,j}+\rho r=\rho \tau ,\qquad \qquad \qquad \qquad \hbox{ on }B\times
(0,\infty ).
\end{array}
\label{1-3}
\end{equation}
In the previous equations:

\begin{quote}
${\bf S}$ and ${\bf f}$ are the stress tensor and body force, respectively;
\newline
${\bf h},\ g$ and $\ell $ are the equilibrated stress vector, intrinsic and
extrinsic equilibrated \newline
body force, respectively; \newline
$\tau ,$ ${\bf q}$ and $r$ are\ the rate at which heat is absorbed for a
unit of volume, the heat \newline
flux vector and the (extrinsic) heat supply, respectively.
\end{quote}

Let ${\bf E}$ be the strain field associated with ${\bf u,\ \bigtriangledown
}\theta ^{t}$ be\ the history up to time t for the thermal gradient and ${%
\overline{\bigtriangledown \theta }}^{t}$ be the integrated history of the
thermal gradient, i. e.
\begin{equation}
E_{ij}=\frac{1}{2}(u_{i,j}+u_{j,i}),  \label{4'}
\end{equation}
\begin{equation}
\theta _{,j}^{t}(s)=\theta _{,j}^{{}}(t-s),\quad \quad \bar{\theta}%
_{,j}^{t}(s)=\int_{t-s}^{t}\theta _{,j}^{{}}(\alpha )d\alpha ,\quad \quad
\forall s\in \lbrack 0,+\infty ).  \label{5}
\end{equation}
As it was shown by Ie\c{s}an in \cite{iesan}, the constitutive equations are
\begin{equation}
\begin{array}{l}
S_{ij}=C_{ijrs}E_{rs}+D_{ijs}\varphi _{,s}+B_{ij}\varphi +M_{ij}\theta , \\%
[4mm]
h_{i}=D_{rsi}E_{rs}+A_{is}\varphi _{,s}+b_{i}\varphi +a_{i}\theta , \\[4mm]
g=-B_{rs}E_{rs}-b_{s}\varphi _{,s}-\xi \varphi -m\theta , \\[4mm]
\rho \tau =-\theta _{0}M_{rs}\dot{E}_{rs}-\theta _{0}a_{s}\dot{\varphi}%
_{,s}-\theta _{0}m\dot{\varphi}+\rho c\dot{\theta},
\end{array}
\label{4}
\end{equation}
where the material coefficients of\ eqs. (\ref{4}) are supposed continuous
and bounded fields on $\bar{B}$ (the set $\bar{B}$ is the closure of $B)$.
Moreover, they satisfy the symmetry relations
\begin{equation}
\begin{array}{l}
C_{ijrs}=C_{rsij}=C_{jirs},\quad D_{ijr}=D_{jir},\quad A_{ij}=A_{ji}, \\%
[4mm]
B_{ij}=B_{ji},\quad M_{ij}=M_{ji}.
\end{array}
\label{4*}
\end{equation}
Further, we consider for the heat flux the following constitutive equation
as suggested by Fabrizio, Lazzari and Munoz Rivera in \cite{Falamu}
\begin{equation}
q_{i}(t)=-\theta _{0}\int_{0}^{\infty }K_{ij}(s)\theta _{,j}^{t}(s)ds,
\label{4bis}
\end{equation}
with the relaxation thermal conductivity ${\bf K}.\ \;$We are assuming $%
K_{ij}$ are continuous and bounded fields on $\bar{B}$; moreover, $%
K_{ij}(s)\in L^{2}(0, \infty )$ and they satisfy the relations

\begin{equation}
K_{ij}(s)=K_{ji}(s),\quad \quad \quad K_{ij}(\infty )=0.  \label{symK}
\end{equation}
By taking into account $\bar{\theta}_{,j}^{t}(0)=0$ and
\begin{equation}
\frac{d\,}{dt}\bar{\theta}_{,j}^{t}(s)=\theta _{,j}(t)-\theta
_{,j}^{t}(s),\quad \quad \quad \frac{d\,}{ds}\bar{\theta}_{,j}^{t}(s)=\theta
_{,j}^{t}(s),  \label{15L}
\end{equation}
the constitutive equation (\ref{4bis}) can be written as

\begin{equation}
q_{i}(t)=\theta _{0}\int_{0}^{\infty }\dot{K}_{ij}(s)\bar{\theta}%
_{,j}^{t}(s)ds.  \label{7}
\end{equation}
In what follows, we study the restriction imposed by the fundamental laws of
thermodynamics in terms of cyclic processes. With this aim, we recall the
laws of the thermodynamics as considered in \cite{Falamu,4,6S}:\bigskip

{\bf First Law of Thermodynamics: }{\it For every cyclic process the
following equality holds}
\begin{equation}
\oint \left\{ \rho \tau (t)+S_{rs}(t)\dot{E}_{rs}(t)+h_{s}(t)\dot{\varphi}%
_{,s}(t)-g(t)\dot{\varphi}(t)\right\} dt=0.\qquad  \label{26}
\end{equation}
\bigskip

{\bf Second Law of Thermodynamics: }{\it For every cyclic process the
following equality holds}
\begin{equation}
\frac{1}{\theta _{0}^{2}}\oint \left\{ \rho \tau (t)(\theta _{0}-\theta
(t))+q_{s}(t)\theta _{,s}(t)\right\} dt\leq 0,  \label{23'}
\end{equation}
{\it and the equality sign holds if and only if the process is reversible. }

\bigskip The relations (\ref{4bis}, \ref{23'}) imply

\begin{equation}
\int_{0}^{d}\left( \int_{0}^{\infty }K_{ij}(s)\theta _{,j}^{t}(s)ds\right)
\theta _{,i}(t)dt\geq 0  \label{23''}
\end{equation}
for any cycle on $[0,\ d)$, and the equality sign holds if and only if $%
\bigtriangledown \theta ^{t}\ $is a constant history.

If we put $\theta _{,i}^{{}}(t)=\bar{k}_{i}\cos (\omega t)+\tilde{k}_{i}\sin
(\omega t)$ into the relation (\ref{23''}) with $d=\frac{2\pi }{\omega }$,
then we obtain
\begin{equation}
\int_{0}^{\infty }(\bar{k}_{i}K_{ij}(s)\bar{k}_{j}+\tilde{k}_{i}K_{ij}(s)%
\tilde{k}_{j})\cos (\omega s)ds+\int_{0}^{\infty }(\tilde{k}_{i}K_{ij}(s)%
\bar{k}_{j}-\bar{k}_{i}K_{ij}(s)\tilde{k}_{j})\sin (\omega s)ds\geq 0,
\label{23*}
\end{equation}
where $\omega $ is strictly positive and $\bar{k}_{i},$ $\tilde{k}_{i}$ are
the continuous functions on $\bar{B}.\ $

Now, we introduce the Fourier transform, sine and cosine transforms of
function $f\in L^{2}(-\infty ,\ \infty )$, i. e.
\begin{equation}
\begin{array}{l}
\displaystyle f^{F}(\omega )=\int_{-\infty}^{\infty }f(\xi )\exp (-i\omega
\xi )d\xi ,\quad \\
\displaystyle f^{S}(\omega )=\int_{0}^{\infty }f(\xi )\sin (\omega \xi )d\xi
,\quad f^{{C}}(\omega )=\int_{0}^{\infty }f(\xi )\cos (\omega \xi )d\xi .
\end{array}
\label{19}
\end{equation}
We remark that the Fourier inversion formula imply
\begin{equation}
f(\xi )=\sqrt{\frac{2}{\pi }}\int_{0}^{\infty }f^{S}(\xi )\sin (\omega \xi
)d\omega ;  \label{four}
\end{equation}
and the Plancherel's theorem of the Fourier transform gives
\begin{equation}
\int_{-\infty }^{\infty }f(\xi )g(\xi )d\xi =\sqrt{\frac{1}{2\pi }}%
\int_{-\infty }^{\infty }f^{F}(\xi )g^{F\,\ast }(\xi )d\omega ,
\label{planch}
\end{equation}
with $f,\ g\in L^{2}($-$\infty ,\ \infty )$ and with $g_{F}^{\ast }$ is the
complex conjugate of $g_{F}.\ $

We can easily prove by eqs. (\ref{15L}, \ref{19}) that
\begin{equation}
\dot{\bar{\theta}_{,i}^{t{C}}}=-\omega \bar{\theta}_{,i}^{t{S}},\qquad
\qquad \dot{\bar{\theta}_{,i}^{t{S}}}=\frac{1}{\omega }\theta
_{,i}(t)+\omega \bar{\theta}_{,i}^{t{C}}.  \label{19bisbis+}
\end{equation}
By putting $\bar{k}_{i}=\tilde{k}_{i}$=$k_{i}$ in the relation (\ref{23*})
as Chirita and Lazzari in \cite{5}, we obtain
\begin{equation}
k_{i}K_{ij}^{C}k_{j}>0,\qquad \qquad \forall {\bf k\equiv (}%
k_{1},k_{2},k_{3})\neq {\bf 0,}\qquad \forall \omega >0.  \label{32L}
\end{equation}
Thus, the relations (\ref{19bisbis+}, \ref{32L}) imply that ${\bf \dot{K}}%
^{S}$\ is a negative definite tensor.

By using eqs. (\ref{7}, \ref{four}, \ref{planch}), we have
\begin{equation}
q_{i}(t)=\sqrt{\frac{2}{\pi }}\theta _{0}\int_{0}^{\infty }\dot{K}%
_{ij}^{S}(\omega )\bar{\theta}_{,j}^{t\,{S}}(\omega )d\omega ,  \label{22}
\end{equation}
and
\begin{equation}
\dot{K}_{ij}(\xi )=\sqrt{\frac{2}{\pi }}\int_{0}^{\infty }\dot{K}%
_{ij}^{S}(\omega )\sin (\xi \omega )d\omega .  \label{23}
\end{equation}
If we integrate eq. (\ref{23}) with the respect $\xi $\ and we take into
account the Riemann Lebesgue lemma and the hypotheses $K_{ij}(\infty )=0$,
then it follows
\begin{equation}
K_{ij}(0)=-\sqrt{\frac{2}{\pi }}\int_{0}^{\infty }\frac{1}{\omega }\dot{K}%
_{ij}^{S}(\omega )d\omega .  \label{25}
\end{equation}
The relations (\ref{symK}, \ref{25}) and the negative definiteness of the
tensor ${\bf \dot{K}}^{S}$\ imply that ${\bf K}(0)$ is a symmetric and
positive definite tensor, i.e.
\begin{equation}
k_{i}K_{ij}(0)k_{j}>0,\qquad \qquad \forall {\bf k\equiv (}%
k_{1},k_{2},k_{3})\neq {\bf 0,\ }\qquad \forall \omega >0.
\end{equation}
Consequently, it follows that
\begin{equation}
K_{ij}(0)k_{i}k_{j}\leq K_{M}k_{i}k_{i},\qquad \qquad \forall {\bf k\ =}%
\{k_{1},k_{2},k_{3}\}{\bf ,\ }  \label{31}
\end{equation}
where $K_{M}({\bf x)}>0$ is the largest characteristic eigenvalue of ${\bf K}%
({\bf x,\ }0).\ $

In what follows, we will assume that the bulk mass density $\rho ,$ the
equilibrated inertia $\chi $ and the constant heat $c$ are strictly
positive, continuous and bounded fields on $\bar{B}$, so that
\begin{equation}
0<\ \rho _{0}\equiv \inf_{{\bf x\in }\bar{B}}\rho ,\qquad 0<\chi _{0}\equiv
\inf_{{\bf x\in }\bar{B}}\chi ,\qquad 0<c_{0}\equiv \inf_{{\bf x\in }\bar{B}%
}c.  \label{7bis}
\end{equation}
Moreover, we suppose that ${\bf K}({\bf x,\ }0)$ is continuous and bounded
on $\bar{B}{\it .}$ By setting
\begin{equation}
K_{0}\equiv \sup_{{\bf x\in }\bar{B}}\frac{\theta _{0}K_{M}}{c},  \label{32}
\end{equation}
we obtain
\begin{equation}
K_{ij}({\bf x,\ }0)k_{i}k_{j}\leq K_{0}\frac{c}{\theta _{0}}k_{i}k_{i}\qquad
\qquad \forall {\bf k\ =}\left\{ k_{1},k_{2},k_{3}\right\} \neq {\bf 0{\it %
.\ }}  \label{42L}
\end{equation}
Now, we introduce the vector space ${\cal V}$ of all vector fields of the
form
\[
{\bf \tilde{F}}\equiv \{F_{ij},\sqrt{\chi }\pi _{i},\psi \},\qquad \qquad %
\hbox{ with}\qquad F_{ij}=F_{ji}.
\]
For any ${\bf \tilde{F}},{\bf \bar{F}}\in {\cal V}$, we define the inner
product and the magnitude by
\begin{equation}
{\bf \tilde{F}}\cdot {\bf \bar{F}}=F_{ij}\bar{F}_{ij}^{{}}+\chi \pi _{i}\bar{%
\pi}_{i}+\psi \bar{\psi},\qquad \left| {\bf \tilde{F}}\right|
^{2}=F_{ij}F_{ij}+\chi \pi _{i}\pi _{i}+\psi ^{2},  \label{8-9}
\end{equation}
where ${\bf \bar{F}}\equiv \{\bar{F}_{ij},\sqrt{\chi }\bar{\pi}_{i},\bar{\psi%
}\}$. For given ${\bf \tilde{F}}$ $\in {\cal V}$, we introduce the vector
field ${\bf \tilde{S}}({\bf \tilde{F}})\in {\cal V}$ as
\begin{equation}
{\bf \tilde{S}}({\bf \tilde{F}})\equiv \left\{ \tilde{S}_{ij}({\bf \tilde{F}}%
),\sqrt{\chi }\left( \frac{1}{\chi }\tilde{h}_{i}({\bf \tilde{F}})\right) ,-%
\tilde{g}({\bf \tilde{F}})\right\} ,\   \label{10}
\end{equation}
where
\begin{equation}
\begin{array}{l}
\tilde{S}_{ij}({\bf \tilde{F}})=C_{ijrs}F_{rs}+D_{ijs}\pi _{s}+B_{ij}\psi ,
\\[4mm]
\tilde{h}_{i}({\bf \tilde{F}})=D_{rsi}F_{rs}+A_{ij}\pi _{j}+b_{i}\psi , \\%
[4mm]
\tilde{g}({\bf \tilde{F}})=-B_{ij}F_{ij}-b_{i}\pi _{i}-\xi \psi ,\
\end{array}
\label{11}
\end{equation}
and the coefficients are defined in eqs. (\ref{4}). They obey to the
symmetry relations (\ref{4*}).

As in \cite{ScStan,Passarella}, we consider the following bilinear form
\begin{equation}
\begin{array}{l}
2{\cal F}({\bf \tilde{F}},{\bf \bar{F}})=C_{ijrs}F_{ij}\bar{F}_{rs}+\xi \psi
\bar{\psi}+A_{ij}\pi _{i}\bar{\pi}_{j}+B_{ij}(F_{ij}\bar{\psi}+\bar{F}%
_{ij}\psi )+ \\[3mm]
\qquad \qquad +D_{ijs}(F_{ij}\bar{\pi}_{s}+\bar{F}_{ij}\pi _{s})+b_{i}(\psi
\bar{\pi}_{i}+\bar{\psi}\pi _{i}).
\end{array}
\label{FiEE}
\end{equation}
From (\ref{11}, \ref{FiEE}) we have
\begin{equation}
2{\cal F}({\bf \tilde{F}},{\bf \bar{F}})=\tilde{S}_{ij}({\bf \tilde{F}})\bar{%
F}_{ij}+\tilde{h}_{i}({\bf \tilde{F}})\bar{\pi}_{i}-\tilde{g}({\bf \tilde{F}}%
)\bar{\psi}.  \label{1ener}
\end{equation}
Throughout this paper, we will assume that the quadratic form $\tilde{W}$
associated to ${\cal F}$ is positive definite, so that
\begin{equation}
2\tilde{W}({\bf \tilde{F}})=2{\cal F}({\bf \tilde{F}},{\bf \tilde{F}})\leq
\mu _{M}\left( F_{ij}F_{_{ij}}+\stackrel{}{\chi }\pi _{i}\pi _{i}+\psi
^{2}\right) ,  \label{stel2}
\end{equation}
where $\mu _{M}>0.\ $In particular, in the case ${\bf \tilde{E}}\equiv
\{E_{ij},\ \sqrt{\chi }\varphi _{,i},\ \varphi \}$, from (\ref{4}, \ref{11})
we get
\begin{equation}
\begin{array}{l}
S_{ij}=\tilde{S}_{ij}({\bf \tilde{E})}+M_{ij}\theta ,\qquad h_{i}=\tilde{h}%
_{i}({\bf \tilde{E})}+a_{i}\theta ,\qquad g=\tilde{g}({\bf \tilde{E})}%
-m\theta .
\end{array}
\label{somma}
\end{equation}
Taking into account eqs. (\ref{11}, \ref{FiEE}, \ref{stel2}, \ref{somma}),
we deduce

\begin{equation}
\begin{array}{l}
W^{\ast }\equiv \tilde{W}({\bf \tilde{E}})=\frac{1}{2}[(S_{ij}-M_{ij}\theta
)E_{ij}+(h_{i}-a_{i}\theta )\varphi _{,i}-(\ g+m\theta )\varphi ], \\
\displaystyle\dot{W}^{\ast }=S_{ij}\dot{E}_{ij}+h_{i}\dot{\varphi}_{,i}-\ g%
\dot{\varphi}+\frac{\rho \tau \theta }{\theta _{0}}-\frac{\rho \theta }{%
\theta _{0}}\dot{\theta},
\end{array}
\label{14bis}
\end{equation}
and
\begin{equation}
2W^{\ast }\leq \mu _{M}\left( E_{ij}E_{_{ij}}+\stackrel{}{\chi }\varphi
_{,i}\varphi _{,i}+\varphi ^{2}\right) .  \label{stel2*}
\end{equation}
By setting ${\bf \bar{F}=\tilde{S}}({\bf \tilde{E})}$ in the relations (\ref
{1ener},\ref{stel2}), we have
\begin{equation}
2\tilde{W}({\bf \tilde{S}}({\bf \tilde{E})})\leq \mu _{M}\left[ \tilde{S}%
_{ij}({\bf \tilde{E})}\tilde{S}_{ij}({\bf \tilde{E})}+\frac{1}{\chi }\tilde{h%
}_{i}({\bf \tilde{E})}\tilde{h}_{i}({\bf \tilde{E})}+\tilde{g}^{2}({\bf
\tilde{E})}\right] .  \label{bo}
\end{equation}
Therefore, we conclude thanks to (\ref{1ener}, \ref{bo}) and thanks to
Cauchy-Schwarz's inequality that
\[
\begin{array}{l}
\displaystyle\left[ \tilde{S}_{ji}({\bf \tilde{E})}\tilde{S}_{ji}({\bf
\tilde{E})}+\frac{1}{\chi }\tilde{h}_{i}({\bf \tilde{E})}\tilde{h}_{i}({\bf
\tilde{E})}+\tilde{g}^{2}({\bf \tilde{E})}\right] ^{2}=\left[ 2{\cal F}({\bf
\tilde{E}},{\bf \tilde{S}}({\bf \tilde{E})})\right] ^{2}\leq 4W^{\ast }%
\tilde{W}({\bf \tilde{S}}({\bf \tilde{E}}))\leq  \\[3mm]
\displaystyle\qquad \qquad \qquad \qquad \leq 2W^{\ast }\mu _{M}\left[
\tilde{S}_{ji}({\bf \tilde{E})}\tilde{S}_{ji}({\bf \tilde{E})}+\frac{1}{\chi
}\tilde{h}_{i}({\bf \tilde{E})}\tilde{h}_{i}({\bf \tilde{E})}+\tilde{g}^{2}(%
{\bf \tilde{E})}\right] .
\end{array}
\]
Consequently, it follows
\begin{equation}
\displaystyle\left| {\bf \tilde{S}}({\bf \tilde{E})}\right| ^{{\bf 2}}{\bf =}%
\tilde{S}_{ij}({\bf \tilde{E})}\tilde{S}_{ij}({\bf \tilde{E})}+\frac{1}{\chi
}\tilde{h}_{i}({\bf \tilde{E})}\tilde{h}_{i}({\bf \tilde{E})}+\tilde{g}^{2}(%
{\bf \tilde{E})}\leq 2\mu _{M}W^{\ast }.\   \label{16}
\end{equation}
Thanks to the inequality for second-order tensors ${\bf L}$ and ${\bf G}$
\[
(L_{ij}+G_{ij})(L_{ij}+G_{ij})\leq (1+\epsilon )L_{ij}L_{ij}+(1+\frac{1}{%
\epsilon })G_{ij}G_{ij},\qquad \forall \epsilon >0,
\]
and thanks to eqs. (\ref{somma}, \ref{16}), we get
\begin{equation}
\begin{array}{l}
\displaystyle S_{ij}S_{ij}+\frac{1}{\stackrel{}{\chi }}h_{i}h_{i}+g^{2}\leq
(1+\epsilon )\left| {\bf \tilde{S}}({\bf E)}\right| ^{{\bf 2}}+(1+\frac{1}{%
\epsilon })\left\{ M_{ij}M_{ij}+\frac{1}{\stackrel{}{\chi }}%
a_{i}a_{i}+m^{2}\right\} \theta ^{2}\leq  \\[4mm]
\displaystyle\qquad \qquad \qquad \leq (1+\epsilon )2\mu W^{\ast }+(1+\frac{1%
}{\epsilon })M\frac{\rho c}{\theta _{0}}\theta ^{2},\qquad \qquad \qquad
\qquad \qquad \qquad \forall \epsilon >0,
\end{array}
\label{17}
\end{equation}
with
\begin{equation}
\displaystyle\mu \equiv \sup_{{\bf x\in }\bar{B}}\mu _{M},\qquad M\equiv
\sup_{{\bf x\in }\bar{B}}\frac{\theta _{0}}{\rho c}\left( M_{ij}M_{ij}+\frac{%
1}{\chi }a_{i}a_{i}+m^{2}\right) .
\end{equation}
The Laws of thermodynamics imply the existence of the thermodynamical
internal potential energy $e$ and the entropy $\eta $\ such that
\begin{eqnarray}
\displaystyle\rho \dot{e}(t) &=&\rho \tau (t)+S_{ji}(t)\dot{E}%
_{ji}(t)+h_{j}(t)\dot{\varphi}_{,j}(t)-g(t)\dot{\varphi}(t),  \label{33} \\
\displaystyle\rho \dot{\eta}(t) &\geq &\frac{1}{\theta _{0}^{2}}\left\{ \rho
\tau (t)(\theta _{0}-\theta (t)+q_{j}(t)\theta _{,j}(t)\right\} .  \label{34}
\end{eqnarray}
We introduce the pseudofree energy potential
\begin{equation}
\rho \Psi (t)=\rho e(t)-\rho \theta _{0}\eta (t).  \label{35'}
\end{equation}
Moreover, we define the maximal pseudofree potential energy $\Psi _{M}$ such
that
\begin{equation}
\displaystyle\rho \dot{\Psi}_{M}(t)=\frac{\rho \tau (t)\theta (t)}{\theta
_{0}}+S_{ji}(t)\dot{E}_{ji}(t)+h_{j}(t)\dot{\varphi}_{,j}(t)-g(t)\dot{\varphi%
}(t)-\frac{q_{j}(t)\theta _{,j}(t)}{\theta _{0}}.  \label{dotf}
\end{equation}
Then, the relations (\ref{33}-\ref{dotf}) imply
\begin{equation}
\rho \dot{\Psi}(t)\leq \rho \dot{\Psi}_{M}(t).  \label{35}
\end{equation}
By taking into account the relations (\ref{4}, \ref{22}, \ref{14bis}$_{2}$),
a maximal pseudofree energy potential is

\begin{equation}
\begin{array}{l}
\displaystyle\rho \Psi _{M}(t)=W^{\ast }(t)+\frac{\rho c\theta ^{2}(t)}{%
2\theta _{0}}-\frac{1}{\pi }\int_{0}^{\infty }\omega \{\dot{K}_{ij}^{S}({\bf %
x},\omega )\bar{\theta}_{,i}^{t{S}}({\bf x},\omega )\bar{\theta}_{,j}^{t{S}}(%
{\bf x},\omega )+ \\
\displaystyle\quad \quad \quad \quad \quad +\dot{K}_{ij}^{S}({\bf x},\omega )%
\bar{\theta}_{,i}^{t{C}}({\bf x},\omega )\bar{\theta}_{,j}^{t{C}}({\bf x}%
,\omega )\}d\omega .
\end{array}
\label{36}
\end{equation}
Since $\tilde{W}$ is a positive definite quadratic form, $\rho ,$ $\chi $\
and $c$ are strictly positive, ${\bf \dot{K}}^{S}$ is the negative definite
tensor, we deduce that the functional $\Psi _{M}$ is a norm. As in \cite{5},
we can prove by eq. (\ref{22}) that
\begin{eqnarray}
\displaystyle\left| {\bf q}(t)\right| ^{2} &=&\frac{2\theta _{0}}{\pi }%
\int_{0}^{\infty }q_{i}(t)\dot{K}_{ij}^{S}({\bf x},\omega )\bar{\theta}%
_{,j}^{t{S}}({\bf x},\omega )d\omega \leq  \label{38} \\
\displaystyle &\leq &\theta _{0}\left( -\frac{2}{\pi }\int_{0}^{\infty }%
\frac{1}{\omega }q_{i}(t)q_{j}(t)\dot{K}_{ij}^{S}(\omega )d\omega \right)
^{1/2}\left( -\frac{2}{\pi }\int_{0}^{\infty }\omega \dot{K}_{ij}^{S}(\omega
)\bar{\theta}_{,i}^{t{C}}(\omega )\bar{\theta}_{,j}^{t{S}}(\omega )d\omega
\right) ^{1/2}.  \nonumber
\end{eqnarray}
The relations (\ref{25}, \ref{32}, \ref{42L}, \ref{36}, \ref{38}) imply
\begin{equation}
\left| {\bf q}(t)\right| ^{2}\leq K_{0}\theta _{0}c\left\{ 2\rho \Psi
_{M}(t)-2W^{\ast }(t)-\frac{\rho c\theta ^{2}(t)}{\theta _{0}}\right\} .
\label{39}
\end{equation}

\section{A time-weighted surface power measure}

Throughout this work by an admissible process we mean an ordered array ${\bf %
\pi }\equiv \lbrack {\bf u}${\bf , }${\bf E,\ S,\ }\varphi {\bf ,\ \gamma ,}$
\newline
${\bf h,\ }\theta ,\ {\bf k,\ q}]$ with the following properties\\[2mm]
i. \thinspace \thinspace \thinspace $u_{i},\ \varphi \in C^{2,2}(\bar{B}%
\times \lbrack 0,\ +\infty )),\ \;\theta \in C^{1,1}(\bar{B}\times \lbrack
0,\ +\infty ))$; \\[2mm]
ii. \thinspace \thinspace\ $E_{ij}=E_{ji},\ \;\gamma _{i}{\bf =}\varphi
_{,i},\ \;k_{i}=\bar{\theta}_{,i}^{t}\in C^{0,1}(\bar{B}\times \lbrack 0,\
+\infty ))$; \\[2mm]
iii. \thinspace\ $S_{ij}=S_{ji},\ \;h_{i},\ q_{i}\in C^{1,0}(\bar{B}\times
\lbrack 0,\ +\infty )),\ $\\[2mm]
and which meets the equations of motion (\ref{1-3}), the geometrical
equations (\ref{4'}, \ref{5}), the constitutive equations (\ref{4}, \ref
{4bis}) and the following initial conditions
\begin{equation}
u_{i}(0)=u_{i}^{0},\quad \quad \dot{u}_{i}(0)=\dot{u}_{i}^{0},\quad \quad
\varphi (0)=\varphi ^{0},\quad \quad \dot{\varphi}(0)=\dot{\varphi}%
^{0},\quad \quad \theta (0)=\vartheta _{0}.  \label{inizio}
\end{equation}
Let ${\bf s}$, $h$ and $q$ be the surface tractions$,$ the surface
equilibrated stress and the heat flux, so
\begin{equation}
s_{i}(t)=S_{ji}(t)n_{j},\qquad h(t)=h_{j}(t)n_{j},\qquad q(t)=q_{j}(t)n_{j}.
\label{3bis}
\end{equation}
We denote by $\Gamma \equiv \lbrack f_{i},\ s_{i},\ u_{i}^{0},\ \dot{u}%
_{i}^{0},\ l,\ h,\ \varphi ^{0},\ \dot{\varphi}^{0},\ r,\ q,\ \vartheta
_{0},\ \theta _{,j}^{0}]$ the external data and assume that all functions
are prescribed continuous functions.

Now, we introduce the support ${\hat{D}}_{T}$ of the external data $\Gamma $
and the body supplies on the time interval $[0,\ T]$, i. e. the set of all $%
{\bf x}\in {\overline{B}}$ such that:

(1) if ${\bf x}\in B$, then
\[
u_{i}^{0}\neq 0\mbox{ or }\dot{u}_{i}^{0}\neq 0\mbox{ or
}\varphi ^{0}\neq 0\mbox{ or }\dot{\varphi}^{0}\neq 0\mbox{ or
}\vartheta _{0}\neq 0
\]
or
\begin{equation}
\theta _{,i}^{0}(s)\neq 0\ \mbox{ for some }s\in (-\infty ,0]
\end{equation}
or
\[
f_{i}(s)\neq 0\ \mbox{ or }\ \ell (s)\neq 0\ \mbox{ or
}\ r(s)\neq 0\ \mbox{ for some }s\in \lbrack 0,T];
\]

(2) if ${\bf x}\in {\partial B}$, then
\[
s_{i}(s)\dot{u}_{i}(s)\neq 0\mbox{ or }h(s)\dot{\varphi}(s)\neq 0\mbox{ or }%
q(s)\theta (s)\neq 0\mbox{ for some }s\in \lbrack 0,T].
\]
In what follows, we will assume that ${\hat{D}}_{T}$ is a bounded set.

We introduce a nonempty set ${\hat{D}}_{T}^{\ast }$ such that ${\hat{D}}%
_{T}\subset {\hat{D}}_{T}^{\ast }\subset {\bar{B}}$ and

(a) if ${\hat{D}}_{T}\cap B\neq \emptyset ,$ then we choose ${\hat{D}}%
_{T}^{\ast }$ to be the smallest bounded regular region in $\bar{B}$ that
includes ${\hat{D}}_{T}$; in particular, we set ${\hat{D}}_{T}^{\ast }={\hat{%
D}}_{T}$ if ${\hat{D}}_{T}$ it is a regular region;

(b) if $\emptyset \neq {\hat{D}}_{T}\subset {\partial B}$, then we choose ${%
\hat{D}}_{T}^{\ast }$ to be the smallest regular subsurface of ${\partial B}$
that includes ${\hat{D}}_{T}$; in particular, we set ${\hat{D}}_{T}^{\ast }={%
\hat{D}}_{T}$ if ${\hat{D}}_{T}$ is a regular subsurface of ${\partial B}$;

(c) if ${\hat{D}}_{T}=\emptyset $, then we choose ${\hat{D}}_{T}^{\ast }$ to
be an arbitrary nonempty regular subsurface of $\partial B$.

On this basis, we introduce the set $D_{r},$ by
\[
D_{r}=\{{\bf x}\in {\overline{B}}:{\hat{D}}_{T}^{\ast }\cap {\overline{%
\Sigma (r)}}\neq \emptyset \},
\]
where $\Sigma (r)$ is the open ball with radius $r$ and center at ${\bf x}$.

Further, we shall use the notation $B_{r}$ for the part of $B$ contained in $%
B\setminus D_{r}$ and we set $B(r_{1},\ r_{2})=B_{r_{2}}\setminus B_{r_{1}}$%
, $r_{1}\geq r_{2}$. We denote by $S_{r}$ the subsurface of $\partial B_{r}$
contained into the inner of $B$ and whose outward unit normal vector is
forwarded to the exterior of $D_{r}$. We can observe that the data are null
on $B_{r},\ S_{r}$.

We define the following time-weighted surface power function $I(r,\ t)$
\begin{equation}
\displaystyle I(r,t)=-\int_{0}^{t}\int_{S_{r}}e^{-\sigma s}[s_{i}(s)\dot{u}%
_{i}(s)+h(s)\dot{\varphi}(s)-\frac{1}{\theta _{0}}q(s)\theta (s)]dads.\
\label{40}
\end{equation}
for a fixed positive parameter $\sigma $\ and for any $r\geq 0,\ t\in
\lbrack 0,\ T]$

The following theorems establish a set of properties for the surface power
function $I$. These theorems will be useful in the study of the spatial
behaviour of the thermoelastic processes.

{\bf Lemma 1: }{\it Let $\pi $ be a thermoelastic process and} {\it $%
\widehat{D}_{T}$ be the bounded support of the external data }$\Gamma ${\it %
\ on the time interval }$[0,\ T].$ {\it Moreover, let }${\it I}${\it $(r,\
t) $ be the time-weighted surface power function associated with $\pi $ and }%
${\cal K}${\it \ be the kinetic energy defined by}
\begin{equation}
{\cal K}{\it =}\frac{1}{2}{\it \bigl(\rho \dot{u}}_{i}{\it \dot{u}}_{i}{\it %
+\rho }\stackrel{}{\chi }{\it \dot{\varphi}}^{2}{\it \bigr)}.  \label{cine}
\end{equation}
{\it \ Under the hypotheses of the Section 2, it follows}

{\it (I) for }${\it 0\leq r}_{2}{\it \leq r}_{1}$%
\begin{equation}
\begin{array}{c}
I(r_{1},t)\ -\ I(r_{2},t)\ =\ -\displaystyle\int_{B(r_{1},r_{2})}{\it e}%
^{-\sigma t}[{\cal K}(t)+\rho \Psi _{M}(t)]{\it dv-} \\
-\sigma \displaystyle\int_{0}^{t}\int_{B(r_{1},r_{2})}{\it e}^{-\sigma s}%
{\it \,}\left[ {\cal K}(s)+\rho \Psi _{M}(s)\right] {\it \ dv\,ds;}
\end{array}
\label{41}
\end{equation}

{\it (II) $I(r,\ t)$ is a continuous differentiable function and
\begin{eqnarray}
\displaystyle\frac{\partial }{\partial r}I(r,t)\  &=&\
-\int_{S_{r}}e^{-\sigma t}\left[ {\cal K}(t)+\rho \Psi _{M}(t)\right] \ da-
\label{42} \\
\displaystyle &&-\sigma \int_{0}^{t}\int_{S_{r}}e^{-\sigma s}\left[ {\cal K}%
(s)+\rho \Psi _{M}(s)\right] \ dads,  \nonumber
\end{eqnarray}
and
\begin{equation}
\displaystyle\frac{\partial }{\partial t}I(r,t)\ =\ -\int_{S_{r}}e^{-\sigma
t}[s_{i}(t)\dot{u}_{i}(t)+h(t)\dot{\varphi}(t)-\frac{1}{\theta _{0}}%
q(t)\theta (t)]da;  \label{43}
\end{equation}
}

{\it (III) for each fixed $t\in \lbrack 0,\ T]$, $I(r,t)$ is a
non--increasing function with respect to $r$. }\bigskip

{\bf Proof.} By using the divergence theorem, by eqs. (\ref{1-3}-\ref{4},
\ref{4*}, \ref{3bis}, \ref{36}, \ref{11}) and thanks to the definitions of $%
\widehat{D}_{T},\ B_{r},\ S_{r},$ we have for $0\leq r_{2}\leq r_{1}$
\begin{equation}
\begin{array}{c}
\displaystyle I(r_{1},t)\ -\ I(r_{2},t)\ =-\int_{0}^{t}\int_{\partial
B(r_{1},r_{2})}e^{-\sigma s}[s_{i}(s)\dot{u}_{i}(s)+h(s)\dot{\varphi}(s)-%
\frac{1}{\theta _{0}}\,q(s)\theta (s)]da\,ds= \\
\displaystyle\qquad \ =-\int_{0}^{t}\int_{B(r_{1},r_{2})}e^{-\sigma
s}[S_{ji,j}(s)\dot{u}_{i}(s)\ +h_{j,j}(s)\dot{\varphi}(s)-\ \frac{1}{\theta
_{0}}\theta (s)q_{i,i}(s)\ +\ S_{ij}(s)\dot{E}_{ij}(s)+ \\[4mm]
\displaystyle\qquad +h_{j}(s)\dot{\varphi}_{,j}(s)-\ \frac{1}{\theta _{0}}%
q_{j}(s)\theta _{,j}(s)]dv\,ds\
=-\int_{0}^{t}\int_{B(r_{1},r_{2})}e^{-\sigma s}\frac{\partial }{\partial s}%
\,\left[ {\cal K}(s)+\rho \Psi _{M}(s)\right] \ dv\,ds.
\end{array}
\label{47}
\end{equation}
\smallskip\

From  (\ref{47}) we get eq. (\ref{41}).

The relation (\ref{41}) implies eq. (\ref{42}),\ taking into account the
concept of derivative as incremental ratio and performing the limit $%
r_{2}\rightarrow r_{1};$ while, the eq. (\ref{43}) comes from the definition
of $I(r,\ t).\ $

The property (III) is obtained from\ (I) and $\rho ,$ $\chi >$0, and by
considering that $\Psi _{M}$\ is a norm. $\blacksquare $

{\bf Lemma 2: }{\it Under hypotheses of Lemma 1, the surface power function $%
I(r,\ t)$\ associated with }$\pi ${\it \ satisfies the following
first--order differential inequalities\smallskip
\begin{equation}
\displaystyle\left| \frac{\partial }{\partial t}I(r,t)\ \right| +\ \zeta
\frac{\partial }{\partial r}I(r,t)\ \leq \ 0,\quad  \label{44}
\end{equation}
}
\begin{equation}
{\it \displaystyle}\frac{{\it \sigma }}{{\it \zeta }}\left| I(r,t)\ \right|
{\it +\ }\frac{\partial }{\partial r}{\it I(r,\ t)\ \leq \ 0,\ }  \label{44+}
\end{equation}
{\it where
\begin{equation}
\zeta =\ \displaystyle\sqrt{\frac{(1+\varepsilon _{0})\mu }{\ \rho _{0}}},
\label{45}
\end{equation}
\smallskip and
\begin{equation}
\displaystyle1+\varepsilon _{0}=\frac{1}{2}\ +\ \frac{(K_{0}+M)}{2\mu }\ +%
\sqrt{\left( -\frac{1}{2}\ +\ \frac{(K_{0}+M)}{2\mu }\right) ^{2}+\frac{M}{%
\mu }}.  \label{46}
\end{equation}
}

{\bf Proof.} By using the Schwarz's inequality and the arithmetic--geometric
mean inequality and the relations (\ref{7bis}, \ref{17}) and (\ref{39}), it
follows that
\begin{equation}
\begin{array}{l}
\displaystyle\left| s_{i}(t)\dot{u}_{i}(t)+h(t)\dot{\varphi}(t)-\frac{1}{%
\theta _{0}}q(t)\theta (t)\right| \ \leq \ \frac{1}{2}\Bigl[\ \frac{%
\varepsilon _{1}}{\ \rho _{0}}\rho \dot{u}_{i}(t)\dot{u}_{i}(t)\ +\frac{1}{%
\varepsilon _{1}}\,S_{ij}(t)S_{ij}(t)+\  \\[4mm]
\qquad \displaystyle+\ \frac{\varepsilon _{1}}{\ \rho _{0}}\rho \chi \dot{%
\varphi}^{2}(t)+\frac{1}{\ \varepsilon _{1}}\,\frac{h_{j}(t)h_{j}(t)}{\chi }%
+\ \frac{\varepsilon _{2}}{\rho _{0}}\frac{\rho c\theta ^{2}(t)}{\theta _{0}}%
+\ \frac{1}{\varepsilon _{2}}\,\frac{q_{j}(t)q_{j}(t)}{c\,\theta _{0}}\ \Bigr%
]\leq  \\[4mm]
\qquad \displaystyle\leq \frac{\varepsilon _{1}}{\ \rho _{0}}{\cal K}(t)+%
\frac{1}{2}\ \Bigl[\frac{1}{\varepsilon _{1}}\left( (1+\epsilon )2\mu
W^{\ast }(t)+(1+\frac{1}{\epsilon })M\frac{\rho c\theta ^{2}(t)}{\theta _{0}}%
\right) + \\[4mm]
\qquad \displaystyle+\ \frac{\varepsilon _{2}}{\ \rho _{0}}\frac{c\rho
\theta ^{2}(t)}{\theta _{0}}+\ \frac{1}{\varepsilon _{2}}\,K_{0}\left( 2\rho
\Psi _{M}(t)-2W^{\ast }(t)-\frac{\rho c\theta ^{2}(t)}{\theta _{0}}\right) %
\Bigr]\ \leq  \\[4mm]
\qquad \displaystyle\leq \frac{\varepsilon _{1}}{\ \rho _{0}}{\cal K}(t)+\ %
\Bigl[\frac{K_{0}}{\varepsilon _{2}}\,\rho \Psi _{M}(t)+\left( \frac{1}{%
\varepsilon _{1}}(1+\epsilon )\mu -\frac{1}{\varepsilon _{2}}\,K_{0}\right)
W^{\ast }(t)+ \\[4mm]
\qquad \displaystyle+\frac{1}{2}\left( \frac{1}{\varepsilon _{1}}(1+\frac{1}{%
\epsilon })M+\frac{\varepsilon _{2}}{\ \rho _{0}}-\frac{K_{0}}{\varepsilon
_{2}}\,\right) \frac{\rho c\theta ^{2}(t)}{\theta _{0}}\Bigr],
\end{array}
\label{48}
\end{equation}
for every $\varepsilon _{1}>0,\ \varepsilon _{2}>0,\ \varepsilon >0.$ We
choose $\varepsilon ,\ \varepsilon _{1},\ \varepsilon _{2}$ such that
\[
\frac{\varepsilon _{1}}{\rho _{0}}=\frac{\ K_{0}}{\varepsilon _{2}},\qquad
\frac{1}{\varepsilon _{1}}(1+\epsilon )\mu -\frac{K_{0}}{\varepsilon _{2}}%
=0\,,\qquad \frac{1}{\varepsilon _{1}}(1+\frac{1}{\epsilon })M+\frac{%
\varepsilon _{2}}{\ \rho _{0}}-\frac{K_{0}}{\varepsilon _{2}}=0;
\]
consequently, we obtain
\[
\varepsilon _{1}=\ \rho _{0}\zeta ,\qquad \qquad \varepsilon _{2}=\frac{%
\,K_{0}}{\zeta },
\]
where $\zeta $ is given by (\ref{45}) and $\varepsilon _{0}$ is given by (%
\ref{46}). In other words, $\varepsilon _{0}$ is the positive root of the
algebraic equation

\begin{equation}
\displaystyle\varepsilon ^{2}-\ \varepsilon (-1\ +\ \frac{K_{0}+M}{\mu })\
-\ \frac{M}{\mu }\ =\ 0.  \label{46bis}
\end{equation}
Thus, the relations (\ref{40}, \ref{43}, \ref{48}) imply

\begin{eqnarray}
\displaystyle\left| \frac{\partial }{\partial t}I(r,t)\right| \ &\leq &\
\zeta \int_{S_{r}}\ e^{-\sigma t}\left[ {\cal K}(t)+\rho \Psi _{M}(t)\right]
da,  \label{dtI} \\
\displaystyle\sigma \left| I(r,t)\right| \ &\leq &\sigma \ \zeta
\int_{0}^{t}\int_{S_{r}}\ e^{-\sigma s}\left[ {\cal K}(s)+\rho \Psi _{M}(s)%
\right] dads.  \label{Idif}
\end{eqnarray}
By eqs. (\ref{42}, \ref{dtI}, \ref{Idif}), the inequalities (\ref{44}, \ref
{44+}) are reached. $\blacksquare $\bigskip

{\bf Lemma 3:}{\it \ Under hypotheses of Lemma 1, the surface power function
}$I(r,\ t)${\it \ associated with }$\pi $ {\it is equal to}
\begin{equation}
\begin{array}{l}
\displaystyle{\it I(r,\ t)=\int_{B_{r}}\ e^{-\sigma t}}\left[ {\cal K}%
(t)+\rho \Psi _{M}(t)\right] \ {\it dv+} \\
\displaystyle\ {\it +\sigma }\int_{0}^{t}\int_{B_{r}}{\it e}^{-\sigma s}%
\left[ {\cal K}(s)+\rho \Psi _{M}(s)\right] {\it \ dvds\geq 0,}
\end{array}
\label{62}
\end{equation}

{\it for }$r\geq 0,t\in \lbrack 0,\ T].\,$

{\bf Proof.} If $B$ is a bounded body, then the variable $r$ ranges on $[0,\
L]$, where
\[
\displaystyle L={max}_{{\bf x}\in \bar{B}}\{{min}_{{\bf y}\in {\widehat{D}%
_{T}^{\ast }}}\,|{\bf x}-{\bf y}|\}<\infty .
\]

From the definition of $\widehat{D}_{T}$ and $I(r,\ t),$ we have $I(L,\ t)=0,
$ so that we obtain eq. (\ref{62}) with the help of eq. (\ref{41}).

On the other hand, if $B$ is an unbounded body, then, the variable $r$
ranges on $[0,\ \infty )$. Fixed a\ pair $(r_{0},\ t_{0})$ in the plane $%
(r,\ t),$ such that $r_{0}\geq \zeta |t-t_{0}|,$ $t_{0}\in \lbrack 0,\ T],$
the inequality (\ref{44}) implies that
\begin{eqnarray}
\displaystyle\frac{d}{dt}\,[I(r_{0}+\zeta (t-t_{0}),t)\,] &\leq &0;
\label{64} \\
\displaystyle\frac{d}{dt}\,[I(r_{0}-\zeta (t-t_{0}),t)\,] &\geq &\ 0.
\label{66}
\end{eqnarray}
Therefore, we have
\begin{eqnarray}
I(r_{0}+\zeta (t-t_{0}),t)\, &\leq &I(r_{0}+\zeta (t^{^{\prime
}}-t_{0}),t^{^{\prime }})\qquad \hbox{ with }t\geq t^{^{\prime }},
\label{64'} \\
I(r_{0}-\zeta (t^{^{\prime }}-t_{0}),t^{^{\prime }}) &\geq &I(r_{0}-\zeta
(t-t_{0}),t).\qquad \hbox{ with }t\geq t^{^{\prime }}.  \label{67}
\end{eqnarray}
For $t=t_{0}$ and $t^{^{\prime }}=0,$ the relations (\ref{40}, \ref{64'},
\ref{67}) imply
\begin{eqnarray}
I(r_{0},t_{0}) &\leq &I(r_{0}-\zeta t_{0},0)=0,  \label{65} \\[4mm]
0 &=&I(r_{0}+\zeta t_{0},0)\geq I(r_{0},t_{0}),  \label{68}
\end{eqnarray}
and so
\begin{equation}
I(r_{0},t_{0})=0.  \label{F59}
\end{equation}
For $r_{0}\rightarrow \infty $ and, consequently, for any $t_{0}\in \lbrack
0,\ T]$, eq. (\ref{F59}) becomes
\[
I(\infty ,t_{0})=\lim_{r_{0}\rightarrow \infty }I(r_{0},t_{0})=0.
\]
From eq. (\ref{41}) and for $r_{1}\rightarrow \infty $ we deduce
\begin{equation}
\begin{array}{l}
\displaystyle I(r,t)=I(r,t)\ -\ I(\infty ,t)\ =\int_{B_{r}}\ e^{-\sigma t}%
\left[ {\cal K}(t)+\rho \Psi _{M}(t)\right] \ dv+ \\
+\sigma \int_{0}^{t}\int_{B_{r}}e^{-\sigma s}\left[ {\cal K}(s)+\rho \Psi
_{M}(s)\right] \ dvds.\ \blacksquare
\end{array}
\label{inf}
\end{equation}
\bigskip

By the properties of the surface power function $I(r,\ t)$ established in
Lemma 1-3$,$ we obtain a complete description of the spatial behaviour of
the elastic process in question outside of the support of the data.

{\bf Theorem 1:}{\it \ Under hypotheses of Lemma 1, for each fixed $t\in
\lbrack 0,\ T]$ we have the following results:}

1) {\it Domain of influence}
\begin{equation}
{\it I(r,\ t)=\ 0,\ \qquad \qquad \hbox{for }r\geq \zeta t.\ }  \label{69}
\end{equation}

{\it 2) Spatial decay}
\begin{equation}
{\it I(r,\ t)\ \leq \ e}^{{\it -}\frac{\sigma }{\zeta }r}{\it I(0,\ t),\
\qquad \qquad \hbox{for }r\leq \zeta t.\ }  \label{2.38}
\end{equation}
{\it \ }

{\bf Proof. }By putting $r_{0}=0,$ $t_{0}=0$ in (\ref{64}), it follows
\[
\displaystyle\frac{dI(\zeta t,t)}{dt}\leq 0,
\]
and so
\begin{equation}
I(\zeta t,t)\leq I(0,0)=0.  \label{70}
\end{equation}
From Lemma 1 and Lemma 3, we have
\begin{equation}
0\leq I(r,t)\leq I(\zeta t,t),{\it \qquad \qquad }\forall r\geq \zeta t.
\label{70bis}
\end{equation}
Therefore, by the inequalities (\ref{62}, \ref{70}, \ref{70bis}) we obtain
the equation\ (\ref{69}).

On the other hand, the inequalities (\ref{44+}, \ref{62}) imply
\begin{equation}
\frac{\partial }{\partial r}\,[{e}^{{\frac{\sigma }{\zeta }r}%
}\,I(r,t)\,]\leq 0,{\it \qquad \qquad }\hbox{for }r\leq \zeta t,
\end{equation}
so that we arrive to (\ref{2.38}). $\blacksquare $

We remark that the relations (\ref{36}, \ref{62}, \ref{69}) give
\begin{eqnarray*}
\displaystyle0 &=&I(r,t)=\int_{B_{r}}e^{-\sigma t}\{{\cal K}(t)+[W^{\ast
}(t)+\frac{\rho c\theta ^{2}(t)}{2\theta _{0}}- \\
&&-\frac{1}{\pi }\int_{0}^{\infty }\omega \{\dot{K}_{ij}^{S}(\omega )\bar{%
\theta}_{,i}^{t{S}}(\omega )\bar{\theta}_{,j}^{t{S}}(\omega )+\dot{K}%
_{ij}^{S}(\omega )\bar{\theta}_{,i}^{t{C}}(\omega )\bar{\theta}_{,j}^{t{C}%
}(\omega )\}d\omega ]\}dv,\quad \hbox{for }r\geq \zeta t.
\end{eqnarray*}
Taking into account that $\rho ,$ $\chi $ and $c$ are strictly positive
functions, that $\tilde{W}$ is a positive definite quadratic form and that $%
{\bf \dot{K}}^{S}$\ is negative definite tensor, we have
\begin{equation}
\dot{u}_{i}=0,\quad \dot{\varphi}=0,\quad \theta =0\qquad \qquad \hbox{on }%
B_{r}\times \lbrack 0,T].  \label{dominiobis}
\end{equation}
Since the external data are null on $B_{r},$ the relation (\ref{dominiobis})
yields

\begin{equation}
u_{i}=0,\quad \varphi =0,\quad \theta =0\qquad \qquad \hbox{on }B\backslash
D_{r}\times \lbrack 0,T]\qquad \hbox{for }r\geq \zeta t.  \label{gutin}
\end{equation}

Now, we establish the domain of influence of the external given data at time
$T,$ according to Gurtin \cite{6}. In fact, we show, by putting $t=T$ and $%
r=\zeta T$ in relation (\ref{gutin}), that on $[0,T]$ the external given
data have no effect on points outside of $D_{\zeta T}$.

{\bf Theorem 2}: {\it Under hypotheses of Lemma 1, it follows}
\begin{equation}
u_{i}=0,\quad \varphi =0,\quad \theta =0\qquad \qquad \hbox{on }B\backslash
D_{\zeta T}\times \lbrack 0,T].  \label{dom}
\end{equation}

As an immediate consequence of Theorem 1, we show the following uniqueness
result valid for a bounded or unbounded body:

{\bf Theorem 3}: {\it Under hypotheses of Lemma 1, there exists at most one
solution for the boundary-initial-value problem.} \smallskip

{\bf Proof. }Thanks to the linearity of the problem, we have only to show
that the null data imply null solution. Let ${\bf \tilde{U}}=\left\{ \tilde{u%
}_{i},\tilde{\varphi},\tilde{\theta}\right\} $ be a solution of the problem (%
\ref{1-3}{\it ,} \ref{4bis}, \ref{3bis}, \ref{inizio}{\it )} corresponding
to null data. In this case, for each $T\in (0,\ +\infty )$ we have $\widehat{%
D}_{T}=\emptyset $ and $I(r,\ t)=0$. Then, we can conclude that
\[
\tilde{u}_{i}=0,\qquad \tilde{\varphi}=0,\qquad \tilde{\theta}=0\qquad
\qquad \hbox{
on }B\times \lbrack 0,+\infty ).
\]

\section{An alternative surface measure}

In the above section, we have assumed that the parameter $\sigma $ is
strictly positive. We can avoid this restriction by considering a measure of
the type
\begin{equation}
\displaystyle P(r,t)=-\int_{0}^{t}\int_{S_{r}}[s_{i}(s)\dot{u}_{i}(s)+h(s)%
\dot{\varphi}(s)-\frac{1}{\theta _{0}}q(s)\theta (s)]dads,\   \label{enr40}
\end{equation}
for $r\geq 0,\ t\in \lbrack 0,\ T]$. Following the procedure developed in
the previous sections, we can easily prove that $P(r,\ t)$ satisfies
\begin{equation}
\displaystyle\frac{\partial }{\partial r}P(r,t)\ \leq \ 0,\quad \quad \quad
\quad \left| \frac{\partial }{\partial t}P(r,t)\ \right| +\ \zeta \frac{%
\partial }{\partial r}P(r,t)\ \leq 0.  \label{**}
\end{equation}
Moreover, we can deduce that $P(r,\ t)$ is equal to the total energy
associated to\ $\pi $\ on $B_{r}$, i.e.
\begin{equation}
P(r,t)=\int_{B_{r}}\ \left[ {\cal K}(t)+\rho \Psi _{M}(t)\right] \;dv\geq 0%
{\it ,}  \label{enr63}
\end{equation}
and
\begin{equation}
P(r,t)\ =0,\qquad \qquad \qquad \hbox{for }r\geq \zeta t.  \label{enrdom}
\end{equation}
On the basis of these results, we establish the following spatial decay
result

{\bf Theorem 4:}{\it \ Let $\pi $ be a thermoelastic process}, {\it $%
\widehat{D}_{T}$ be the bounded support of the external data }$\Gamma ${\it %
\ on the time interval }$[0,\ T]$ {\it and\ }${\it P}${\it $(r,\ t)$ be the
surface measure associated with $\pi $. Under hypotheses of Section 2, for
each fixed $t\in \lbrack 0,\ T]$ we have
\begin{equation}
Q(r,t)\ \leq \ \left( 1-\frac{r}{\zeta t}\right) Q(0,t),\qquad \qquad \hbox
{for }r\leq \zeta t,  \label{Q}
\end{equation}
\ where}
\begin{equation}
{\it Q(r,\ t)=}\int_{0}^{t}{\it \ P(r,\ \alpha )\;d\alpha =}%
\int_{0}^{t}\int_{B_{r}}{\it \ }\left[ {\cal K}({\it \alpha })+{\it \rho
\Psi }_{M}{\it (\alpha })\right] {\it \;dvd\alpha .\ }  \label{enr62}
\end{equation}
{\bf Proof}{\it .} By (\ref{enrdom}, \ref{enr62}) we have
\[
Q(r^{\prime },t^{\prime })=\int_{\frac{r^{\prime }}{\zeta }}^{t^{\prime }}\
P(r^{\prime },\alpha )d\alpha ,
\]
for any choose $r^{\prime }>0$, $t^{\prime }\in (0,\ T]$ so that $r^{\prime
}\leq \zeta t^{\prime }.$ Now, we consider the following changed variable
\begin{equation}
\alpha =\left( 1-\frac{r^{\prime }}{\zeta t^{\prime }}\right) s+\frac{%
r^{\prime }}{\zeta };  \label{enr73}
\end{equation}
clearly, $\alpha \geq s.$ We get
\begin{equation}
Q(r^{\prime },t^{\prime })=\left( 1-\frac{r^{\prime }}{\zeta t^{\prime }}%
\right) \int_{0}^{t^{\prime }}\ P(r^{\prime },\left( 1-\frac{r^{\prime }}{%
\zeta t^{\prime }}\right) s+\frac{r^{\prime }}{\zeta })ds.  \label{b}
\end{equation}
On the other hand, as in the Lemma 3, we can prove that
\begin{equation}
\frac{d}{dt}P(r_{0}+\zeta (t-t_{0}),t)\ \leq 0.  \label{enr70}
\end{equation}
Thus, by putting $r_{0}=r^{\prime }$, $t_{0}=\alpha $ into (\ref{enr70})$,$
we reach the result
\begin{equation}
\ P(r^{\prime },\alpha )\,\leq P(r^{\prime }+\zeta (s-\alpha ),s)=P\left(
\frac{r^{\prime }s}{t^{\prime }},s\right) \hbox{
with }\alpha \geq s.  \label{enr74}
\end{equation}
\ The first inequality of (\ref{**}) implies that\ $P(r,t)$\ is
non-increasing function of $r$, such that

\begin{equation}
P\left( \frac{r^{\prime }s}{t^{\prime }},s\right) \leq \ P(0,s)\qquad
\hbox{
with }s\in \lbrack 0,t^{\prime }].  \label{++ll}
\end{equation}
For $r^{\prime }\leq \zeta t^{\prime }$\ and $t^{\prime }\in \lbrack 0,\ T],$
we deduce by the inequality (\ref{enr74}, \ref{++ll}) that
\begin{equation}
P\left( r^{\prime },\left( 1-\frac{r^{\prime }}{\zeta t^{\prime }}\right) s+%
\frac{r^{\prime }}{\zeta }\right) \leq \ P(0,t^{\prime }).  \label{a}
\end{equation}
Finally, the relations (\ref{enr62}, \ref{b}, \ref{a}) yield to the
inequality (\ref{Q}). $\blacksquare $

However, we have to outline here that the decay rate characteristic to the
estimate (\ref{Q}) is lower than the established one in the above section.

\end{document}